# Edu-Edition Spreadsheet Competency Framework

Maria Csernoch, Piroska Biró
University of Debrecen, Debrecen, Hungary Kassaiút 26. 4028
csernoch.maria@inf.unideb.hu, biro.piroska@inf.unideb.hu

**ABSTRACT**

*Based on the Spreadsheet Competency Framework for finance professionals, in the present paper we introduce the Edu-Edition of the Spreadsheet Competency Framework ($E^2SCF$). We claim that building spreadsheet competences should start in education, as early as possible, and this process is a lot more effective if support arrives from expert teachers. The main feature of $E^2SCF$ is high mathability computer-supported real world problem solving. This approach is based on – from the very beginning of training – a two-directional knowledge transfer, data and error analysis and handling, and the programming aspect of spreadsheets. Based on these features, $E^2SCF$ is set up for basic and general users to build up firm spreadsheet knowledge and to develop transferable problem solving skills and competences.*

## 1. INTRODUCTION

### 1.1. Twenty Principles and the Spreadsheet Competency Framework

Two fundamental documents dealing with spreadsheets were introduced and published in the last two EuSpRIG conferences: "Twenty principles for good spreadsheet practice" and "Spreadsheet competency framework – A structure for classifying spreadsheet ability in finance professionals" (SCF) in 2015 and 2016, respectively. However, like any pioneer work, these theses are in need of several revisions until a consensus is reached in the community. The second edition of "Twenty principles for good spreadsheet practice" was already published and welcomed in 2016. In the present paper we provide the details of our remarks, comments, and suggestions for reconsidering the SCF, and present the Edu-Edition of the Spreadsheet Competency Framework ($E^2SCF$).

Our suggestions for the revision are based on the role, importance, and influence of teaching spreadsheets, focusing on knowledge transfer between the different subfields of computer sciences and informatics (ICT for short) and other traditional sciences– closely related to the teachers' beliefs in the incremental nature of science –, the influence and role of expert teachers (Hatie, 2003; Chen et al., 2015; Csernoch, 2017), computational thinking (Wing, 2006), schema-construction (Skemp, 1971), and the educational aspect of end-user studies and training in general.

### 1.2. About the Edu-Edition of the Spreadsheet Competency Framework

The primary feature of this edition is that general end-user training and education have been inserted into the spreadsheet framework. We formulated the educational version on the basis of the original Spreadsheet Competency Framework. In this context we focus on the triad of (1) computational thinking, (2) knowledge transfer, and (3) computer problem solving (CT–KT–CPS), how these phenomena strengthen each other, and can be placed in spreadsheet environments. The widely accepted "classical" approach is that "[t]he typical spreadsheet user just wants to use a power tool, and the craft comes later."

However, both our research in spreadsheet training/education and research in education-effectiveness have proved that those who are self-taught or have learned spreadsheets separately from other birotical software ('bureautique' or office applications) or further ICT environments, or focus on the use of the spreadsheet interfaces and/or tools, do not develop the "deep understanding of the science". This means that their knowledge is limited in the sense that it is not open to further input, cannot be transferred to other environments, and is hardly open to further development. With this approach the "sunk-cost" fallacy also takes its toll. It means that once end-users and even teachers make some progress armed with "the belief in the fixed nature of science" they are reluctant to try and learn approaches which might require computational thinking – a method more demanding at the beginning than "click here and there" training – but more effective in the long run. To re-train those who learned the less effective methods is highly demanding, both mentally and financially. Instead, we suggest an early start with the CT–KT–CPS triad, which can widen the participants' perspectives and their understanding of the science, as well as building firm fundamental spreadsheet knowledge.

**1.3. "Classical" methods vs. Functional Modelling**

The other framework into which the CT–KT–CPS triad can be fitted is Functional Modelling (Hubwieser2004; Schneider 2004, 2005). Based on this model, functional programming – programming in general – is highly supported, and spreadsheets are used and taught as a simplified language. Sprego – Spreadsheet Lego – is placed in this framework, with its further simplifications, which match the requirements of introductory programming.

Both the "classical" and the CT–KT–CPS approaches have their advantages and disadvantages. The advantage of the "classical" approach is that training focuses on special features, and trainees can be framed into interfaces to gain this special knowledge in a reasonably short period of time. However, as has already been mentioned, the disadvantage of this method is that this knowledge is difficult to transfer, does not support schema construction, and consequently, knowledge is not built up in long term memory (Csernoch & Biró,2013, 2014;Biró & Csernoch, 2014). In short, end-users forget what they learn, and they have to be provided with training on the same level several times in order to acquire some reliable knowledge. In the long run this approach is time consuming and highly demanding, considering both human and technical resources.
The disadvantage of the CT–KT–CPS triad is that it takes time. However, this is normal, since it develops fundamental skill – computational thinking (Wing, 2006) – which is as important in the digital era as the classical skills: reading, writing, and arithmetic – known as the 3Rs. In the long run, developing basic skills are more important than short term interface and tool management.

The development of basic skills should be started as early as possible. However, the approach can also work with adults. Our research has proved that young adults in tertiary education at first are reluctant to accept this novel approach (the "sunk-cost" fallacy in action, again). However, we have found that their results improve significantly after they make progress in developing their computational thinking and algorithmic skills.

Furthermore, in general education we cannot only focus on those students who are oriented towards engineering, scientific research, medicine, etc. With our model we would like to reach an even wider public, even those who have no mathematical background beyond the four basic operators. This is one of the reasons that we fundamentally changed the introductory methods, and focus on text-based problems and tasks, instead of numerical examples. We have found that using non-mathematical problems in our training is a lot more motivating and they are as good as other problems for developing the algorithmic skills of the students. Furthermore, non-mathematical problems support computational thinking, computer problem solving, and human-computer interaction better than pure mathematical and financial problems.

## 2. Analysing the "Spreadsheet competency framework"

The SCF is mainly built for the closed community of finance professionals, despite claims to the contrary (SCF, 2016, p.2, p.3; Section 2.1). This community is closed in two senses:(1) the group of finance professionals is only a small, well defined proportion of spreadsheet users with clearly distinguished and distinguishable skills, and (2) the SCF does not leave space for the further development of these end-users. Beyond that, the document does not consider spreadsheet management as part of the greater ICT community (Section 2.1), which would be crucial in the acceptance of end-user computing (Panko, 2013, 2015; Panko & Port, 2013;Kadijevich, 2013).In general, this approach leads to a situation in which the SCF and spreadsheet management are secluded, both from the digital and the traditional sciences.

### 2.1. The SCF: the "Preface" and "About this framework" Chapters

The preface of the SCF clearly presents two of the greatest problems related to spreadsheet knowledge and competences (SCF, 2016, p.2): (1) "Spreadsheet skills are often learned ad hoc – almost two-thirds of Excel Community users are self-taught.", (2) "Many users are unaware of their own true competency. Novices are generally overconfident; experts tend to sell themselves short."

Furthermore, four levels of users are defined and described in this introductory section (SCF, 2016, p.3):basic users (BU),general users (GU),creators (CR),and developers (DE). The original SCF claims that "…while the levels are designed with a finance function in mind, their content is largely applicable to any person that uses spreadsheets in their job." However, we have found that the SCF is effectively narrowed down to financial purposes, indicating a closed type of spreadsheet usage, instead of emphasizing the openness of the software and the wide range of possible goals.

### 2.2. The SCF: Chapter "The framework specification"

The core of the SCF is "The framework specification". This chapter consists of a table of 1 + 4 columns (SCF, 2016, p.5–7). In the first, non-titled, column a rather random selection of tools, skills, competences, and activities are listed and named as items. Neither the selection criteria of these items nor their presented order are known. At Level 1 ten major categories are named and each main category contains further subcategories. The other four columns are reserved to separate the levels of users.

A cell is identified by the section of an item and a level. The character of the cell indicates whether that item is required or not at that level. For each item and level three options are available:

- if the cell is empty, the item is not required at that level,
- a white circle indicates the core items (in the present paper the ● character),
- a black circle indicates the beneficial items (in the present paper the ○ character).

The white and black circles play a crucial role in the framework, since they indicate the requirements clearly. In the case of the core items, this is knowledge the users must have, while in the case of the beneficial items, it is knowledge they should have.

However, according to the document, the phenomena which are substituted by the expression "item" are interchangeable, regardless of their original content. It is also clear that more tools are listed than skills. Furthermore, competences, which should be the focus of the work, according to its title, cannot be clearly recognized in the document.

We also suggest a change in the characters used for the core and the beneficial items. The filled circle suggests more requirements; it is more emphatic than the empty circle. In the present paper the original white circle is substituted with the ●, while the black with the ◎ character.

**Selected examples of misjudged items**

In the following, we list various items and their levels, which we consider to be misjudged and/or misplaced in the SCF. Due to the size restriction of the present paper, we primarily focus on those items which are the subject of knowledge transfer and which can be moved to basic and/or general levels in the W$^2$SCF. This schema focuses on the development of computational thinking, which is a basic skill which all professionals should have in the digital era, and on which they can base the special requirements of their professions. (The items are presented with their main and subcategories with the → character between them, while the levels are indicated with the –, ●, ◎ characters.)

Item: Data analysis→Excel tables→ Use data stored in an Excel table
Level:      ●–                    ◎ GU, CR, DE

It is not clear what content and competences are connected to the expression "Use data stored in an Excel table". Spreadsheet management is about using data stored in these tables. Consequently, several different competences on the different levels can be assigned to this item. However, to go into details, considering all the competences in connection with "use data" is far beyond the scope of the present paper.

Item: Data analysis→Excel tables→Use Excel tables to manage data
Level:      ●–                    ◎ CR, DE

Similar to the previous item, all the levels would use spreadsheet tables to manage data. For basic users, these include simple or well protected tables, and for higher levels more complicated tables with more demanding tasks to solve.

Item: Development and problem solving→ Break down and research problem
Level:      ● DE                  ◎CR

In our context "problem solving" means solving tasks, formulating and answering questions based on the available data or on the requirement(s) of the user. Beyond that, in our framework computer tools – both hardware and software – play a fundamental role; ultimately they provide the tools to solve the formulated problems. "This usage has basically two forms: in some cases we use existing functions and methods provided by a system, and we apply these tools to solve the problems. Another possibility is, if we, based on existing means of the system, develop new programs and functions for solving new problems." (Baranyi & Gilányi, 2013), i.e. low- and high-mathability problem solving, respectively (Biró & Csernoch, 2015a, 2015b).

Handling errors is only one element of general problem solving. This special section is part of the discussion and debugging process of high-mathability problem solving.

Even for the simplest problem, and consequently for the lowest level of users, problem solving skills should be a fundamental requirement. Without analysing the problem users should not start using any software. Aimless wandering around the interface usually leads to unreliable output.

---

Item: Development and problem solving→Trace errors in spreadsheet they build
Level:      ● CR, DE                    ○–

---

Discussion and error recognition should be part of the activities at any level .Just like the previous item, this must also be present at the lowest level. Even the simplest data recording activities cannot be managed safely without tracing errors. Users have to be trained from the very beginning that they should be aware of all their activities and their consequences.

**Examples of miscategorised items**

In the following we list items which are referred to as spreadsheet competences, but which are not specifically spreadsheet but general ICT or other traditional subject-based knowledge.

---

Item: Basic skills→ Access and save files, Read and enter data, Set up and printing
Level:      ●BU, GU, CR, DE             ○–

---

All these three items are general ICT skills and should be transferred to spreadsheets; consequently, they should not be listed as spreadsheet competency items.

---

Item: Design and best practice→ File naming and version control
Level:      ● DE                        ○GU, CR

---

"File naming and version control" is closely related to the three previously mentioned items. Similar to them, handling files is a general ICT skill, and moreover, "version control" is extremely software-specific. For a deeper understanding, instead of "version control" the more general "conversion with Save as" would be preferable as a core item on all the levels. This option includes "version control", and in addition, knowledge transfer would serve users' interests better. Furthermore, "conversion with Save as" would lead users to file conversions, types, extensions, and file naming in general, which is a reversal knowledge transfer (discussed in Section 3.1), although it is beyond the scope of the present paper.

One further remark in connection with "version control" is that it is extremely time consuming and rather challenging to follow all the changes which MS applies to Excel. With 500+ functions, and their – not uncommonly – ambiguous lists of arguments and descriptions, erroneous and constant-based examples (Section 2.3;Csernoch 2014, 2017), there is no chance for effective and error free document handling. Beyond this, the older versions of Help are extremely difficult to access, even on the Internet, which makes "version control" even more challenging. The managing of "version control" was one of the reasons which led us to compose a simplified, version-independent, spreadsheet-based programming language, Sprego (Csernoch, 2014; Csernoch & Biró, 2015a, 2015b, 2015c, 2016a, 2016b; Biró & Csernoch, 2015a, 2015b; Section 3.2).

---

Item: Efficiency of use → Shortcuts → Navigation shortcuts, Find and replace
Level:      ● GU, CR, DE           ○ BU
Item: Efficiency of use →Shortcuts → Copy and paste shortcuts
Level:      ● BU, GU, CR, DE       ○ –

---

"Navigation shortcuts", "Copy and paste shortcuts", "Find and replace" can be transferred from general ICT knowledge.

---

Item: Efficiency of use→Shortcuts→Level 3: Additional shortcuts
Level:      ● –                    ○CR, DE

---

The "Additional shortcuts" expression is so general that it should not be listed in the framework specification table.

---

Item: Formulas → Text formulas
Level:      ● –                    ○ GU, CR, DE

---

According to our ultimate goal, to set up the Edu-Edition of the framework, we must be aware that handling texts serves both beginners and students extremely well. On the one hand, novice users are more comfortable with strings than numbers, especially students of elementary and low high school classes, and users not specialized in mathematics, finance, economics, etc., On the other hand, displaying and handling characters in spreadsheets is well supported and, along with this capability of spreadsheets, text-based functions offer great support for an understanding of the different data types.

**2.3. SCF: Chapter "Explanatory notes to the framework"**

The explanatory section of the SCF (SCF, 2016, p.8–18) is intended to clarify the items of the "framework specification" table and gives reasons for the inclusion of some of the items. The list of items in the "framework specification" makes it clear that the SCF tends to focus on the tools rather than the skills and competences, and this tendency is strengthened in the "explanatory notes" chapter.

The examples selected for the present paper refer to low mathability activities, where the tools are the focus, instead of the problem solving approaches, which is in accordance with the position of problem solving in the "framework specification" and the levels assigned to it (Section 2.2). Problem solving in this framework is handled as something mysterious which is the privilege of a select few.

Basic and general users are guided towards mechanical activities, and the SCF does not require them to do any creative work. However, we argue that any user level would carry out creative work within the context of their range of ability, and the SCF should define these skills and competences instead of requiring the mere use of a piece of software or a software family.

**Examples of arithmetic and logical formulas in the SCF**

Our research and practice proved that handling arithmetic formulas for basic and general users, as is detailed in Section 3.2, is more demanding than working with texts. The case is similar with logical operators, a phenomenon which is unknown at this level, without experience in programming. Novices understand the expression yes/no question, transferred from language studies, better than logical test/logical expression/condition. The reformulation of help expressions would also support novices in understanding programming concepts (Csernoch, 2014, 2017).

The logical formulas– condition-based problem specific functions –, listed in the explanatory section, along with the family of COUNT() functions, are the black-sheep of the spreadsheet family. While the basic IF() function is mandatory (Csernoch, 2014; Csernoch & Biró, 2015a, 2015b, 2015c, 2016a, 2016b; Biró & Csernoch, 2015a, 2015b), the other *IF?() functions are completely unnecessary and loaded with serious restrictions (Csernoch, 2014): they are version dependent; only the AND connection is defined, no formula is allowed in the test, inequality is handled as a string instead of as an operator, using variables is not supported. Furthermore, there are varying lists of arguments, unknown programming concepts in Help descriptions, categorised in different function groups, and a limited number of functions for only a limited number of problems. This latest restriction is completely "non-programming". One of the tools with which MS has tried to make spreadsheets more user-friendly is the introduction of novel functions. However, MS does not recognize that they would never be able to introduce as many functions as necessary to solve all the problems in the world, even though requests arrive from the user voice forum. The user voice forum is not relevant in this context, since we argue against the "classical" ineffective training and usage of spreadsheets, whose main feature is the hundreds of functions. It is obvious that MS developed the *IF?() functions in response to users' demand. However, we proved that these functions never reached the wider public. It was found in our testing that end-users can only handle the COUNTIF() function to some extent: with equality and constant in the condition. They do not know the name of the other *IF?() functions, how to handle inequality, and variables in these functions, even in the COUNTIF().

Instead, we argue that if we train students and end-users to apply the algorithm of these problems they would be a lot more flexible, version independent, and ultimately more effective. The algorithm for substituting problem specific logical functions– COUNT*() and *IF?() functions –is the following: (1) formulating yes/no(s) question, (2) making decisions on the output values of the TRUE, FALSE branches, and (3) applying the instruction to the output values of the previous step (Csernoch, 2014; Csernoch & Biró, 2015a, 2015b, 2015c, 2016a, 2016b; Biró & Csernoch, 2015a, 2015b). With this algorithm only the yes/no questions have to be formulated properly and only one function, the IF(), has to be understand thoroughly. Consequently, all the COUNT*() and *IF?() functions become unnecessary.

There is always a serious debate regarding how deep we need to go into the functional model. For example, do we have to include the AVERAGE() function or not in our basic set? If we go deeper than the AVERAGE() function, we need a method to count the number of elements involved. However, there is no need for the COUNT() function either; it can be replaced with the SUM(IF()) composite function, based on the algorithm detailed above. Beyond that the COUNT() function with all its alterations has become rather confusing for end-users not specialized in maths.

There are two further reasons which do not support extreme abysses in the functional model. One is discussed in the present paper (Section2.2): even calculating an average with the SUM() and the COUNT() combination is too complicated maths for young children. The other reason is that research has proved that students do not have to know the theoretical background to solve maths problems, but can apply the available computer tools effectively (Chmielewska & Gilányi, 2015; Chmielewska et al., 2016). Consequently, we have found more advantages than disadvantages for including the AVERAGE() function in basic and general training.

**Examples of lookup formulas in SCF**

Lookup functions are crucial but problematic in spreadsheets. There are remnants of older versions, which might cause version inconveniences. However, the major problem originates in the HLOOKUP()/VLOOKUP() functions. These two functions are highly supported in spreadsheets forums, even though they are unnecessarily complicated and carry serious restrictions. The MATCH() function would be an alternative solution but both the function's wizard and its Help facility, along with the SCF's explanatory notes (SCF, 2016, p. 12), are incorrect and inconsistent (Csernoch, 2014, 2017).

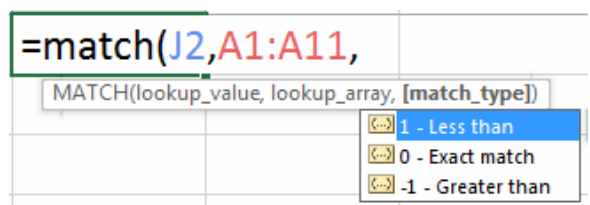

Figure 1. The thumb messages associated with the MATCH() function

The common errors related to the lookup functions can be demonstrated with the available Help facilities of the MATCH() function.

- Level 1 is the thumb message (Fig. 1).The *lookup value* is correct. However, the second argument, the *lookup_ array* is not, since only a one-dimensional array is accepted, so the correct argument should be *lookup_vector*. The third argument is the *match_type*, where the "less than" and the "greater than" expression appear, which are not much help. It would be much better to say that 1, !1, or 0 has to be typed when the values of the vector are in ascending order, descending order, or non-ordered, respectively. Then the thumb message is expanded: "Finds the largest value that is less than or equal to lookup_value. Lookup_array must be placed in ascending order." The message only at the end contains the most important information: "Lookup_array must be placed in ascending order.", however, no average user gets this far.

- Level 2 is the wizard. At this level, the description of the MATCH() function is the following: "Returns the relative position of an item in an array that matches a specified value in a specified order." While the definition of the *lookup_value*: "Lookup_value is the value you use to find the value you want in the array, a number, text, or logical value, or a reference to one of these." There is no end-user who can understand these. It would be easier to say that the "lookup_value is the value you are looking for".

- Level 3 is the official help. This is the best, since it gives e.g. the following description: "The MATCH function searches for a specified item in a range of cells, and then returns the relative position of that item in the range." and the following short definition for the *lookup_value* "The value that you want to match in lookup_array.". We have to go the third level to get some valuable information about the MATCH() function. However, no end-user would do this on regular bases.

In the original SCF in the comparison of the lookup functions the following partially correct description can be read: "Approximate matches with VLOOKUP, HLOOKUP and MATCH will only work if the first column of the table being referred to is sorted correctly."

- In the case of the VLOOKUP() and HLOOKUP() functions only ascending order is allowed, while in the MATCH() function both ascending and descending orders are permitted. This is one of the advantages of the MATCH() function over the VLOOKUP() and HLOOKUP(). Beyond that, all three functions handle non-ordered data.

- For the HLOOKUP() function the first row should be ordered, not the first column. With the MATCH() function it does not matter whether the vector is a row or a column. This is the second advantage of the MATCH() function.

- In the case of the MATCH() function the order of the vector with the matrix does not matter, which is the third advantage of the MATCH() function. Consequently, the "first column/row" restriction is only applies to the VLOOKUP() and HLOOKUP() functions.

Considering all these points, a well-defined algorithm can be formulated to handle vector based items: (1) defining the look-up value, (2) finding the position of the record, and (3) displaying the selected value of the record in question. This algorithm can be coded extremely conveniently with the INDEX(MATCH()) composite function. The first step of this algorithm cannot be defined more precisely, since it is highly dependent on the problem to be solved (Csernoch, 2014).

A further advantage of teaching and using both SEARCH() and MATCH() functions is that this approach provides strong support for knowledge transfer within spreadsheets. The two functions are based on similar algorithms and the output value of both is a position, while the only difference between them is the input data type, with all its consequences.

**Examples of building error-resistant formulas in the SCF**

As is detailed in this paper (Section 3), recognising and handling errors must be a part of the training process from the very beginning. Manual data analysis and recognition of data types would serve novices extremely well. However, general users can handle error outputs with the ISERROR() function, and consequently, with the IF(ISERROR()) composite function. As was mentioned earlier with other groups of functions, there is no need for problem specific functions handling different errors, since this combination would serve all purposes. In this context, where computational thinking is focused on, one further advantage of the simple ISERROR() function is that users are "forced" to figure out the source of the error, which is a very similar process to the selection of the appropriate error function and the understanding of its output.

In the SCF error-examples, similar to the previous ones, the focus is again on technical details and version compatibility, instead of the algorithm of handling errors, along with explanatory error messages and their role in programming.

## 3. The Edu-Edition of the "Spreadsheet competency framework"

The following two hypotheses play an extremely important role in our reasoning; spreadsheet competences should be considered (1) at an earlier stage than the "finance professional" and (2) for a much wider range of users. In general, we argue that developing spreadsheet competences should start in and with education. Taking this into consideration, in what follows, we introduce the "Edu-Edition Spreadsheet Competency Framework" ($E^2SCF$).

In the $E^2SCF$ we argue that developing spreadsheet competences must be started in formal education as early as possible, focusing on real world problem solving with high mathability approaches (Biró & Csernoch, 2015a, 2015b),which support knowledge transfer and offer a more general computer science approach to spreadsheets.

One of the major points of our argument is that problem solving and knowledge transfer should be focused on when considering digital competences. In this sense, problem solving includes the major steps of real world problem solving: (1) analysing the problem, (2) building algorithms, (3) coding, and (4) discussion, debugging (Pólya, 1954). As was mentioned earlier, problem solving should not be the privilege of highly qualified spreadsheet professionals. It should be part of the training and the application of spreadsheets at any level, starting from basic users. Problem solving and programming should not be mystified (Ben-Ari, 2011). Instead, the difficulty level of problems should be defined and matched with the level of users (Skemp, 1971).Similar to real world problem solving, design, data analysis, and error handling should also be present at any level.

Data analysis, especially unplugged and/or semi-automated data analysis, should be introduced as early as possible. End-users have to be trained to recognize data types, the contents, the possible input and output values, and the connection between them. Without this background knowledge and skill, reliable data management is almost impossible.
In the problem solving approach to end-user training and application schemata-construction also plays a crucial role. This is the key concept for knowledge transfer and reliability. If complete, general schemata are available, the knowledge built up earlier, regardless of the environments, can be transferred and applied in novel situations, and these schemata also lead to a reduction in the number of errors in documents, since fast, intuition-based thinking becomes reliable (Csernoch, 2017; Kahneman, 2011).

### 3.1. The benefit of spreadsheet knowledge transfer

**Knowledge transfer**

When working with spreadsheets, a two-directional knowledge transfer is preferable and should be available. Firstly, spreadsheets are open to incoming knowledge and schemata built up in other ICT and traditional environments. Secondly, due to the programming aspect of spreadsheets, knowledge built up in this user-friendly interface can be transferred to more traditional programming environments, to database management, to ICT – data handling and coding, end-user text management, etc. –, and to traditional sciences, primarily mathematics. Finally, due to the simplicity of the spreadsheet

environments– considering the data manipulating interface and the toolbars –and the available high mathability spreadsheet-programing approaches and languages(Booth, 1992;Hubwieser, 2004;Wakeling, 2007;Sestoft, 2011; Csernoch, 2014) – considering the algorithm-building and the coding –,even inexperienced debutant and end-user programmers have the opportunity to focus on the problems, the algorithms, and the related discussions, instead of on the coding details.

Spreadsheets would be an ideal introductory problem solving and programming environment if widely accepted institutions and concepts– such as SCF,ECDL, MS, etc. – focused on this aspect, instead of on the tools and the pure usage itself. In the following we outline E$^2$SCF, which strongly supports knowledge transfer with high mathability problem solving approaches.

### E$^2$SCF for basic and general users

In E$^2$SCF, first we suggest the modification of the structure of the "framework specification" table. A column should be added to the table which considers the general ICT and traditional subject knowledge brought into spreadsheets– MAthematics (MA), Design and Planning (DP), Incremental Nature of Science (IS) (Chen et al., 2015), authentic contents (AC) –,input knowledge for short (IK). We also suggest a change in the order of the Level 1 items and in several cases changes in the subgroups. Since we focus on general education and the two-directional knowledge transfer, our competency table considers only two levels; basic and general users (BU and GU). The competences listed in the table would serve as the basis for further studies in ICT and also as a tool for strengthening concepts in other sciences, especially in mathematics.

We can read in the SCF that "individuals below the basic user level should not be in a position to access an organisation's spreadsheets, as they are unlikely to use them safely and effectively." (SCF, 2016 p. 4) We go one step further and claim that there is a certain ICT background which is required for working safely and effectively in computer related activities. Until this level is reached, individuals should not be allowed to work in digital environments for any company. However, we must also note that working in spreadsheet environments would develop and strengthen users' ICT knowledge and security. A reversal knowledge transfer is possible if spreadsheets are not taught in isolation but as part of the greater ICT community. Consequently, training must also focus on this aspect of knowledge transfer.

Table 1. The framework specification of E$^2$SCF considering the knowledge brought into spreadsheet management through knowledge transfer and spreadsheet competences of basic and general users (BU and GU, respectively)

| Item | Input knowledge | BU | GU |
|---|---|---|---|
| **Problem solving** | | | |
| Breaking down and researching problems | MA, DP, IS, AC | ● | ● |
| Tracing errors in spreadsheets they build | MA, ICT, IS, AC | ● | ● |
| Building error-resistant formulas | MA, ICT | ● | ● |
| Understanding manual vs automatic calculation | ICT, MA, IS, AC | ● | ● |
| Recognizing error messages | ICT, MA, IS, AC | ● | ● |
| Handling data-entering error messages | ICT, MA, AC | ● | ● |
| Handling formula-entering error messages | MA, AC | ● | ● |
| Handling data-driven error messages | AC | | ● |
| Recognizing data types | ICT | ● | ● |
| Analysing data manually | MA, ICT | ● | ● |

| Item | Input knowledge | BU | GU |
|---|---|---|---|
| **Basic ICT skills** | | | |
| Accessing and saving files | ICT | ● | ● |
| Reading and entering data | ICT | ● | ● |
| Manipulating set up and printing | ICT | ● | ● |
| Naming files | ICT | ● | ● |
| Converting files with Save As | ICT | ● | ● |
| Managing find and replace processes | ICT | ● | ● |
| Understanding and applying navigation shortcuts | ICT | ● | ● |
| Understanding and applying copy and move shortcuts | ICT | ● | ● |
| Understanding and applying file management shortcuts | ICT | ● | ● |
| **Design and best practice** | | | |
| Designing layout | DP, ICT, AC | ● | ● |
| Explaining calculations they build | ICT, MA, AC | ● | ● |
| **Formulas** | | | |
| Understanding and applying basic arithmetic | MA | ● | ● |
| Understanding the concept of functions | MA, ICT | ● | ● |
| Calling non-array-based general purpose functions | | ● | ● |
| Understanding and handling vectors | | ● | ● |
| Building vector output array formulas | | ● | ● |
| Building one value output array formulas | | ● | ● |
| Calling array-, error-, and condition-based general purpose functions | | | ● |
| Building 2 and 3-level composite functions | | ● | ● |
| Building multi-level composite functions | | | ● |
| Understanding precedent and dependent cells | | ● | ● |
| **Formatting** | | | |
| Understanding and applying hiding, unhiding, deleting, inserting rows, columns, cells | ICT | ● | ● |
| Understanding and applying grouping, merging | ICT | | ● |
| Understanding and applying regular cell formatting | ICT | ● | ● |

### 3.2. Tools for managing E²SCF

**Array formulas vs. copying and references**

Introducing array formulas at the basic level might seem challenging. However, our testing has proved that even beginners welcome the concept, learn it fast, and apply it safely (Biró&Csernoch,2016a, 2016b;Csernoch & Biró, 2013, 2014). One of the advantages of array formulas (Walkenbach & Wilcox, 2003; Walkenbach & Wilcox, 2003;Walkenbach, 2002, 2010; Csernoch, 2014) is that both copying formulas and using absolute and mixed references are avoidable, which plays a crucial role in spreadsheet security, since these are two of the major sources of spreadsheet errors. Furthermore, using vector-output array formulas, single items cannot be changed, unlike in copying, which makes spreadsheet formulas safer. In educational environments – in a class room – one further advantage of array formulas is that it makes the modification of the formulas faster, in the sense that teachers do not have to check and warn students repeatedly whether they have modified the formula in the first instance nor remind them about copying.

Beyond this, array formulas strongly support the usage of variable vs. constants and single-value array formulas can substitute various problem specific built-in formulas. The other reason that absolute and mixed references are left out from E$^2$SCF is that experiences prove that the phenomenon of references is one of the most difficult concepts in spreadsheets; consequently, basic and general users are not ready for it. In general, if array formulas are built there is no need for absolute and mixed references, at this level.
Considering the handling of array formulas, we have found that it is more natural to use them in tables, with data arranged into fields – vectors – than handling all the cells as individual items. We also tried our method with primary and middle school children and found that with our unplugged tools it is a lot easier to handle a set of data as one object (Biró & Csernoch, 2017). Beyond that, due to the user-friendly environments of spreadsheets, the definition/declaration of a vector is only a selection, which is extremely convenient. We are aware of the debate over the changing of the size of the arrays when we handle vector-outputs. However, there are methods in which the changes in the size can be handled in flexible ways.

One further advantage of defining and using array formulas in spreadsheets is that students can be prepared for programming in imperative languages. The concept of array – vector and matrix – is introduced in an environment where the declaration and definition of array is extremely convenient: only a selection of a range on the graphical interface. Beyond this, the concept of loop is also introduced, since with an array formula repeated activities are carried out on the items of the arrays.

**Functions**

Research and experience have proved that for beginners no more than 12–15 functions or instructions can be taught and used effectively (Hromkovic, 2009;Walkenbach,2002, 2010; Walkenbach & Wilcox, 2003; Wilcox & Walkenbach, 2003). Based on these findings, we defined a dozen functions, entitled Sprego functions (Csernoch, 2014), which serve as the introductory set for basic and general users. It has also been experienced that trainees without any special education in mathematics are better at handling text based and text oriented problems, and consequently the functions handling these problems. Considering all these, the set of Sprego functions consists of

- four text functions– LEN(), LEFT(), RIGHT(), SEARCH(),
- four maths functions – SUM(), AVERAGE(), MIN(), MAX(),
- four functions for handling conditions, arrays, and errors – IF(), MATCH(), INDEX(), ISERROR().
-

According to the problems, the set of Sprego functions can be expanded with further general purpose functions. Our suggestion is the following: SUBSTITUTE(), SMALL(), LARGE(), AND(), OR(), NOT(), INT(), ROUND(), RAND(), OFFSET(), ROW(), COLUMN().

Using only a limited number of functions has the advantage that students can remember them, so knowledge and schemata can be stored in long term memory, and can be called up and activated in problem solving in fast, safe, and effective ways.

One further advantage is that handling other, non-mathematical functions would clarify the concept of function introduced in maths classes. In spreadsheet environments, and especially in Sprego, n-ary and composite functions are introduced, handled, and required for problem solving, something which– currently –is mostly avoided in elementary and high-school mathematics. With opportunities to work with "real" n-ary functions a reversal knowledge transfer is possible, from ICT to mathematics. Handling composite functions in Sprego is similar to traditional programming languages: the decision to build composite functions or use additional variables and/or arrays is always guided and ruled by the requirements and nature of the problem and the programming environment. The creation of the composite functions and embedded structures would also be part of the reversal knowledge transfer. A concept which is hardly mentioned in maths classes, is however, a basic element of knowledge in traditional programming.

**Real world problem solving and the levels of understanding in spreadsheets**

The $E^2SCF$ extensively supports the usage of authentic tables – used in Csernoch, 2014;Csernoch & Biró, 2015a, 2015b and defined in Csernoch & Biró, 2017. In brief, authentic tables contain real data whose content can be selected in accordance with the students' interest, and as such can be highly motivating (Angeli, 2013; Ainley& Pratt, 2005; Cooper & Dunne, 2000) and easily converted into real world situations. Beyond considering the content, with this method, incoming knowledge in the form of handling files is activated in an intensive way. Consequently, authentic tables can provide data which motivate students to use spreadsheets. Research has clearly proved that one of the reasons for failure when teaching spreadsheets is the decontextualized and technocentric teaching methods (Angeli, 2013; Csernoch & Biró, 2016a, 2016b; Mireault, 2016;Csernoch, 2017), a tendency which is recognizable in the original SCF. Finally, motivation and spreadsheet-supported problem solving also play a crucial role in the acceptance of end-user computing, something which does not happen at present (Panko, 2013, 2015, Panko & Port, 2013, Kadijevich, 2013).

To test and evaluate solutions for real world spreadsheet problems we adapted the SOLO categories of understanding (Biggs & Collis, 1982;Lister et al., 2006),originally set up for programming tasks and problems (Biró & Csernoch, 2014). With this method we are able to follow the students' selection of functions, their effectiveness, and their understanding of the formula they selected, and the algorithm they built to solve the problem.

Our tests prove that the usage of the problem specific functions is mainly restricted to formulas holding constants and equality, and one function can only serve one specific problem. Beyond this, our tests clearly show that learning problem specific functions with restricted usage does not support either knowledge transfer or schema-construction. Consequently, those students who are grounded in these functions do not have the ability to generalize; they apply low mathability, ineffective, and erroneous solutions, if they apply any at all.

## 4. Conclusions

The present paper introduces the Edu-Edition of the Spreadsheet Competency Framework (E$^2$SCF). We argue that spreadsheets should be taught and handled as part of the greater ICT world, focusing on two-directional knowledge transfer, data management, the programming aspect of spreadsheets, and computer aided real world problem solving, in general.

Based on the Spreadsheet Competency Framework for finance professionals, we set up the education framework specification for basic and general trainees and users. Compared to the original framework specification, this table contains an additional column which presents the incoming knowledge necessary for faster and more reliable spreadsheet management: ICT bases and problem solving abilities and skills transferred from other sciences and ICT environments. The other major feature of our framework specification is that problem solving is required at any level, starting from novice end-users, which involves real world problem solving based on authentic tables and contents.

Beyond the framework specification table, we provided the essence of Sprego programming which is a supporting tool for E$^2$SCF. With this approach not only can firm schema-based spreadsheet knowledge be built through real world problem solving, but a reversal knowledge transfer is supported, which affects further studies in ICT, especially in programming and data management, and influences other traditional sciences.